\RequirePackage{amsthm}
\documentclass[pdflatex,sn-mathphys-num]{sn-jnl}

\usepackage{graphicx}%
\usepackage{multirow}%
\usepackage{amsmath,amssymb,amsfonts}%
\usepackage{amsthm}%
\usepackage{mathrsfs}%
\usepackage[title]{appendix}%
\usepackage{xcolor}%
\usepackage{textcomp}%
\usepackage{manyfoot}%
\usepackage{booktabs}%
\usepackage{algorithm}%
\usepackage{algorithmicx}%
\usepackage{algpseudocode}%
\usepackage{listings}%

\usepackage{changes}
\definechangesauthor[name={Celi Lou}, color=blue]{lcl}

\theoremstyle{thmstyleone}%
\theoremstyle{thmstyletwo}%

\theoremstyle{thmstylethree}%

\newcommand{\codeURL}{\url{https://github.com/LiQiushiWIUCAS/ETNet}}

\newcommand{\dataURL}
{\url{https://doi.org/10.6084/m9.figshare.30436504.v1}}

\begin{document}

\title[Article Title]{Computational TIRF enables optical sectioning beyond the evanescent field for widefield fluorescence microscopy}


\author[1]{\fnm{Qiushi} \sur{Li}}
\equalcont{These authors contributed equally to this work.}

\author[1,2]{\fnm{Celi} \sur{Lou}} 
\equalcont{These authors contributed equally to this work.}

\author[1,2]{\fnm{Yanfang} \sur{Cheng}}

\author[1]{\fnm{Bilang} \sur{Gong}}

\author[1]{\fnm{Xinlin} \sur{Chen}}

\author[1]{\fnm{Hao} \sur{Chen}}

\author[1,3]{\fnm{Baowan} \sur{Li}}

\author[1]{\fnm{Jieli} \sur{Wang}}

\author[1]{\fnm{Yulin} \sur{Wang}}

\author[1]{\fnm{Sipeng} \sur{Yang}}

\author*[1]{\fnm{Yunqing} \sur{Tang}}\email{tangyunqing@wiucas.ac.cn}

\author*[1]{\fnm{Luru} \sur{Dai}}\email{dai@wiucas.ac.cn}

\affil*[1]{\orgdiv{Wenzhou Key Laboratory of Biomedical Imaging, Wenzhou Institute}, \orgname{University of Chinese Academy of Sciences}, \orgaddress{ \city{Wenzhou}, \postcode{325000}, \state{Zhejiang}, \country{China}}}

\affil[2]{\orgdiv{School of Medicine and Health}, \orgname{Harbin Institute of Technology}, \orgaddress{ \city{Harbin}, \postcode{150001}, \state{Heilongjiang}, \country{China}}}

\affil[3]{\orgdiv{School of Instrument Science and Engineering}, \orgname{Harbin Institute of Technology}, \orgaddress{ \city{Harbin}, \postcode{150001}, \state{Heilongjiang}, \country{China}}}


\abstract{
The resolving ability of widefield fluorescence microscopy is fundamentally limited by out-of-focus background owing to its low axial resolution, particularly for densely labeled biological samples. Although total internal reflection fluorescence (TIRF) microscopy provides strong near-surface sectioning, they are intrinsically restricted to shallow imaging depths. Here we present computational TIRF (cTIRF), a deep learning-based imaging modality that generates TIRF-like sectioned images directly from conventional widefield epifluorescence measurements without any optical modification. By integrating a physics-informed forward model into network training, cTIRF achieves effective background suppression and axial resolution enhancement while maintaining consistency with the measured widefield data. We demonstrate that cTIRF recovers near-surface structures with performance comparable to experimental TIRF, and further enables both single-frame and volumetric sectioned reconstruction in densely labeled  samples where conventional TIRF fails. This work establishes cTIRF as a practical and deployable alternative to hardware-based optical sectioning in fluorescence microscopy, enabled by rapid adaptation to new imaging systems with minimal calibration data.}


\maketitle

\section{Introduction}\label{sec-intro}

Fluorescence microscopy serves as a cornerstone technology in cell biology, enabling visualization of tissue and subcellular structures. However, widefield (WF) fluorescence microscopy suffers from substantial out-of-focus background and limited axial resolution\cite{stephens_light_2003,liu_breaking_2018}, which degrade performance on deep cell and tissue imaging. To address these limitations, both hardware-based microscopy enhancements and computational algorithms were developed.

Hardware-based microscopy techniques have significantly enhanced optical sectioning capability, such as laser-scanning confocal microscopy \cite{stephens_light_2003}, multiphoton microscopy \cite{zipfel_nonlinear_2003,chen_vivo_2021} and light-sheet microscopy \cite{stelzer_light_2021}. Beyond improving sectioning, techniques such as 3D structured illumination microscopy (SIM)\cite{wu_faster_2018,li_three-dimensional_2023,zhang_super-resolution_2025}and Total internal reflection fluorescence (TIRF) microscopy \cite{mattheyses_imaging_2010} further achieve axial super-resolution, breaking the diffraction limit along the optical axis. However, compared to WF microscopy, these hardware-centric approaches faced inherent limitations, including reduced temporal resolution, increased system complexity, and limited sample applicability. In contrast, algorithm-based computational methods offer a powerful alternative to enhance image quality without necessitating additional hardware modifications or costs.

Algorithmic innovations aim to overcome the fundamental trade-offs inherent in hardware-based methods, redefining strategies for optical sectioning and super-resolution. Deconvolution algorithms attempt to enhance both lateral and axial resolution\cite{zhao_sparse_2022,wernersson_deconwolf_2024,sage_deconvolutionlab2_2017,richardson_bayesian-based_1972,lucy_iterative_1974,bertero_introduction_2021}, yet their effectiveness remains constrained by the ill-posed nature of inverse problems. Specifically, the misattribution of out-of-focus background as in-focus signals persists. Additional algorithms have emerged to mitigate this issue, such as Dark sectioning \cite{cao_dark-based_2025} and Wavelet-based background subtraction method \cite{hupfel_wavelet-based_2021}. Beyond classical algorithms, data-driven deep learning methods were proposed to achieve isotropic resolution restoration. Supervised approaches typically rely on synthetic training datasets due to the scarcity of high-resolution ground truth (GT) images\cite{weigert_content-aware_2018,ma_pretraining_2024}. However, real-world applicability of supervised learning is often limited by inaccuracies in the priors used for synthetic data generation\cite{qiao_evaluation_2021,cai_toward_2019}. Self-supervised strategies circumvent the need for high-resolution GT\cite{park_deep_2022,ning_deep_2023}, successfully enabling isotropic restoration. However, their effectiveness requires sufficient z-slices and low background for input image stacks. Among them, emerging cross-modality techniques leverage real-world data to translate low-quality images (e.g., low-resolution/low-contrast) to high-quality modalities\cite{wang_deep_2019,ma_deep_2021,xu_cross-modality_2023,bouchard_resolution_2023,huang_enhancing_2023}. Despite their success, current studies predominantly focus on improving lateral resolution, largely overlooking opportunities for axial resolution enhancement.

Inspired by cross-modality learning concepts, we introduce computational TIRF (cTIRF) as a computational imaging modality that capable of suppressing out-of-focus background and enhancing axial-resolution from a single widefield epifluorescence image. cTIRF achieved a TIRF-comparable imaging quality, while overcoming the intrinsic depth limitation of hardware-based TIRF and enabling imaging throughout whole cells and tissue sections. To facilitate practical deployment, we enable rapid adaptation of cTIRF to new imaging systems through few-shot learning, achieved by integrating physically motivated image degradation modeling into the reconstruction framework. The cTIRF representations are quantitatively validated against experimentally measured TIRF and confocal microscopy. Furthermore, we extend cTIRF to volumetric reconstruction by employing knowledge distillation \cite{wang_knowledge_2022} to train a three-dimensional network that requires only six optical sections, enabling fast and faithful three-dimensional imaging under limited axial sampling. Together, this work establishes cTIRF as a powerful and user-friendly computational imaging modality for enhancing widefield fluorescence microscopy, with broad applicability to biological and biomedical imaging.

\section{Results}\label{sec-result}

\subsection{Computational TIRF enables optical sectioning and axial super-resolution for a single WF image}

\begin{figure}[htbp]
\centering
\includegraphics[width=0.9\textwidth]{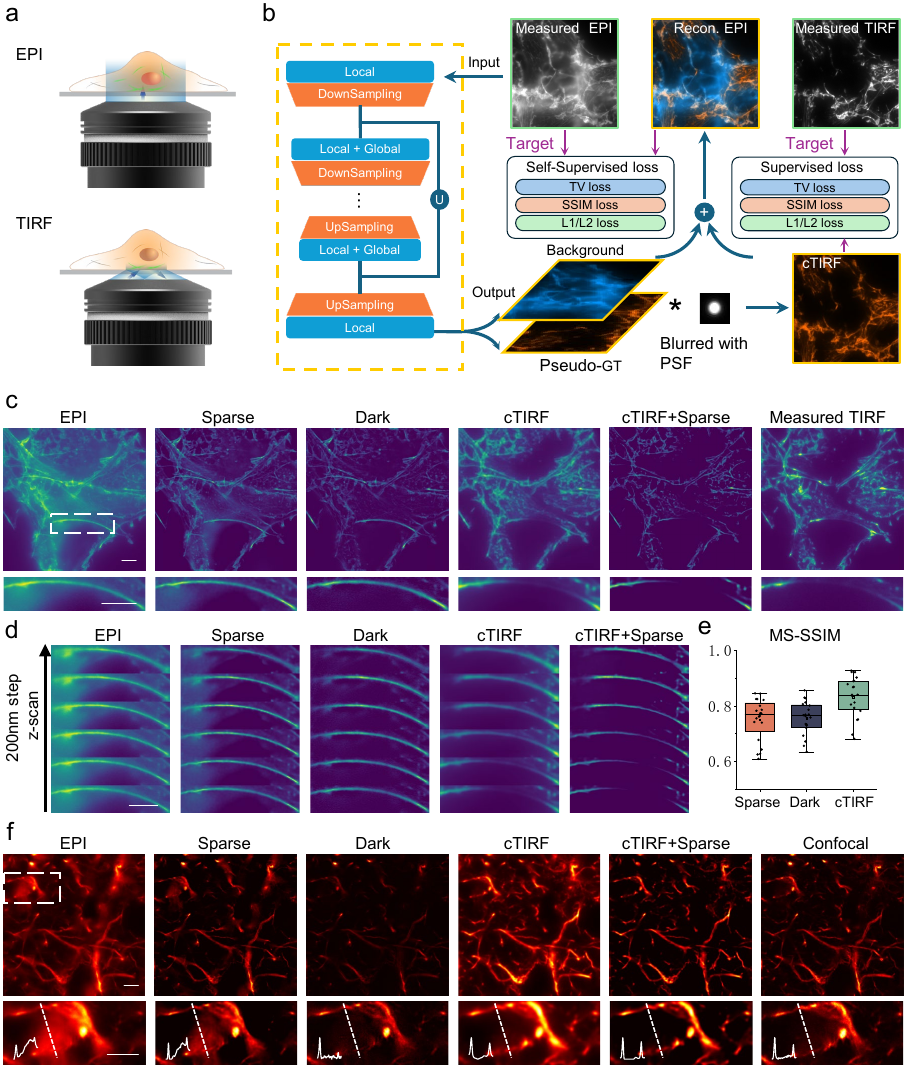}
\caption{\textbf{Overview of computational TIRF for a single WF image.} 
\textbf{a} EPI and TIRF illumination modes. 
\textbf{b} The architecture of ET2dNet. 
\textbf{c} Representative F-actin labeled C6 cell image of EPI(first column), Sparse-Deconvolution reconstruction (second column), Dark Sectioning reconstruction (third column),
cTIRF (fourth column), Sparse-Deconvolution reconstruction of cTIRF (fifth column),  and experimental measured TIRF (sixth column). The magnified view of region in white box is displayed at the bottom.  
\textbf{d} EPI z-scan stack of white dashed box highlighted region in (c), with 200nm step size, and corresponding reconstruction. 
\textbf{e} Average MS-SSIM (to TIRF) of processed images in test dataset, by either Sparse-Deconvolution, Dark Sectioning and cTIRF.  
\textbf{f} Representative images of human brain tissue section stained with an antibody against GFAP,  EPI(first column), Sparse-Deconvolution reconstruction (second column), Dark Sectioning reconstruction (third column),
cTIRF (fourth column), Sparse-Deconvolution reconstruction of cTIRF (fifth column),  and confocal (sixth column). The magnified view of region in white box is displayed at the bottom, with marker line and corresponding intensity profile. 
 Scale bar: 5 $\rm \mu m$.
 }\label{fig-ET2dNet}
\end{figure}

\subsubsection{training 2D cross-modality network using paired EPI-TIRF images}

TIRF microscopy is widely used in biological and biomedical research, exhibiting axial super-resolution, minimal out-of-focus background, and improved signal-to-noise ratio (SNR). It is based on the phenomenon of total internal reflection (as shown in Fig.\ref{fig-ET2dNet} a), where an evanescent wave is generated at the interface and penetrates only a very short distance (typically 50-200 nm) into the sample, exciting fluorophores only in a thin layer near the bottom of the cell-containing coverslip. Hence, TIRF has no ability to image deep fluorophores or complete cell structure. Our purpose was to develop a deep learning-based method, combining the advantages of both EPI (deep illumination depth) and TIRF microscopy (high SNR and axial-resolution).

For deep neural network training, we acquired 881 registered EPI-TIRF image pairs of F-actin-labeled C6 cells using a custom-built EPI-TIRF imaging system (Methods \ref{sec-setup}). Each paired acquisition maintained identical x-y-z positions, with TIRF images in focus. Rigid registration (Methods \ref{sec-registration}) was implemented to compensate for potential stage drift artifacts.

We developed the EPI-TIRF cross-modality network, denoted ET2dNet, with a hybrid architecture. ET2dNet comprised dual branches for self-supervised and supervised learning to generate two outputs, pesudo-GT and background, as shown in Fig.\ref{fig-ET2dNet} b. The in-focus pesudo-GT was convolved with the point spread function (PSF) to generate computational TIRF. For simplification, we define pesudo-GT and computational TIRF (cTIRF) as the output produced by our model; unless otherwise specified, pesudo-GT and cTIRF images are generated using ET2dNet (2D) or ET3dNet (3D reconstruction network, see latter sections). A supervised loss was computed by comparing cTIRF and experimental measured TIRF. Then the cTIRF and background were combined to generate a reconstructed EPI, enabling self-supervised training through comparison with experimental measured EPI. For the network backbone, we implemented EViT-UNet \cite{li_evit-unet_2025}, a state-of-the-art (SOTA) segmentation network. Details are shown in Methods \ref{sec-ETNet}.

The hybrid network architecture drew conceptual inspiration from the classical deconvolution theory\cite{bertero_introduction_2021}, where iterative gradient descent minimizes reconstruction errors through regularization-guided optimization by incorporating a prior. In this deep learning framework, the supervised learning branch encoded biological priors through training on experimental measured EPI-TIRF pairs, and the self-supervised learning branch explicitly modeled the image degradation process via PSF convolution. The generalization capability of the architecture, enhanced by its self-supervised component (as discussed in the latter section), aligns with recent advances in microscopic\cite{qiao_zero-shot_2024} and natural\cite{chen_low-res_2024} image super-resolution.

The EViT-UNet backbone was selected to capture global image features essential for recognizing out-of-focus signals, which typically span large spatial regions in images. While Transformer-based architectures addressed the inherent limitation of CNNs (e.g., standard U-Net\cite{navab_u-net_2015}) in modeling global context\cite{liang_swinir_2021,vaswani_attention_2017}, their high computational cost remained a barrier. EViT-UNet overcame this by integrating efficient self-attention with convolutional operations, preserving global feature extraction capabilities while drastically reducing computational complexity.

\subsubsection{cTIRF shows efficient background subtraction and axial-resolution enhancement}
cTIRF removed background and achieved axial super-resolution. We implemented ET2dNet on the EPI image and compared the output (cTIRF) with the experimental measured TIRF image, as shown in Fig.\ref{fig-ET2dNet} c. Visually, the cTIRF showed minimal background and high contrast. Then we focused on the region highlighted by the white dashed box, where the measured EPI and TIRF images revealed distinct characteristics of the microfilament structure. EPI image showed a continuous microfilament, while the TIRF image depicts a discontinuous structure. Given the shallow penetration depth of TIRF illumination (approximately 50-200 nm) at the bottom in the z-direction, we inferred that the microfilament formed an arch-like structure, with the middle part positioned away from the bottom. This structure served as a typical example to assess the axial-resolution enhancement achieved by cTIRF. We then acquired six EPI images along the z-axis, spaced 200 nm apart, as illustrated in Fig.\ref{fig-ET2dNet} d. Visually, these images appeared nearly indistinguishable. However, the corresponding cTIRFs exhibited distinct differences and collectively reconstruct an arch-like structure, demonstrating the enhanced resolution achieved by cTIRF. The corresponding full image stacks are provided in Supplementary Information Fig. S1.

We compared the performance among cTIRF, SOTA deconvolution (Sparse Deconvolution\cite{zhao_sparse_2022}) and background removal (Dark Sectioning\cite{cao_dark-based_2025}) methods, as shown in Fig.\ref{fig-ET2dNet} c. 
The Sparse Deconvolution gave a sharper reconstruction, while the out-of-focus signal still existed. The Dark Sectioning could efficiently remove the background, but barely enhanced the axial resolution, which was clearly shown by the region of the arch-like structure. It should also be noted that, ET2dNet was not specifically designed to implement deconvolution, but it was compatible with deconvolution methods, as shown in the sixth row in Fig.\ref{fig-ET2dNet} d. By directly implementing Sparse Deconvolution on the cTIRF, both lateral and axial resolution were enhanced. We further examined cTIRF on a test dataset containing 20 images, generating a better average MS-SSIM (compared to Sparse Deconvolution and Dark Sectioning), as shown in Fig.\ref{fig-ET2dNet} e. Additional examples of F-actin labeled U87 cells are shown in Extended Data Fig.1. We also compared other deconvolution and background removal methods, including RL deconvolution\cite{richardson_bayesian-based_1972,lucy_iterative_1974}, Wavelet-based background and noise subtraction (WBNS)\cite{hupfel_wavelet-based_2021}, Background Filtering (BF)\cite{mo_quantitative_2023}, while cTIRF outperformed all these methods, see Extended Data Fig.2. Brief introductions and implementation details of the above algorithms are provided in Supplementary Information Note 1. The details about the combination of cTIRF and Sparse deconvolution are provided in Supplementary Information Note 2.

The axial resolution enhancement achieved by ET2dNet was estimated using image decorrelation analysis \cite{descloux_parameter-free_2019}. An EPI z-stack of an F-actin–labeled C6 cell was processed with ET2dNet (applied independently to each XY slice) to generate a cTIRF stack, and axial resolution was quantified on the corresponding XZ/YZ cross-sections using sectorial decorrelation analysis (Methods \ref{sec-decorrelation}). This analysis indicates an average axial-resolution improvement of approximately threefold.

The performance of cTIRF was further assessed using WF-confocal image pairs of human brain tissue sections stained with an antibody against GFAP (provided by ref.\cite{wernersson_deconwolf_2024}), as shown in Fig.\ref{fig-ET2dNet} f. Note that the WF image was collected using 60X 1.4NA oil objective (with pixel size 107.9nm), which was slightly different from the training dataset using 60X 1.5NA oil objective (with pixel size 97.5nm). In such case, cTIRF could still efficiently remove background, outperforming the Sparse Deconvolution method and Dark Sectioning. As indicated by the region in the white box and the corresponding intensity profile, Sparse Deconvolution failed to suppress background, while Dark Sectioning erroneously eliminated signals. In contrast, cTIRF achieved a background-free result while preserving critical structural information, demonstrating high consistency with confocal GT.

\subsection{Computational TIRF can be rapidly adapted to new imaging systems}

\begin{figure}[H]
\centering
\includegraphics[width=0.9\textwidth]{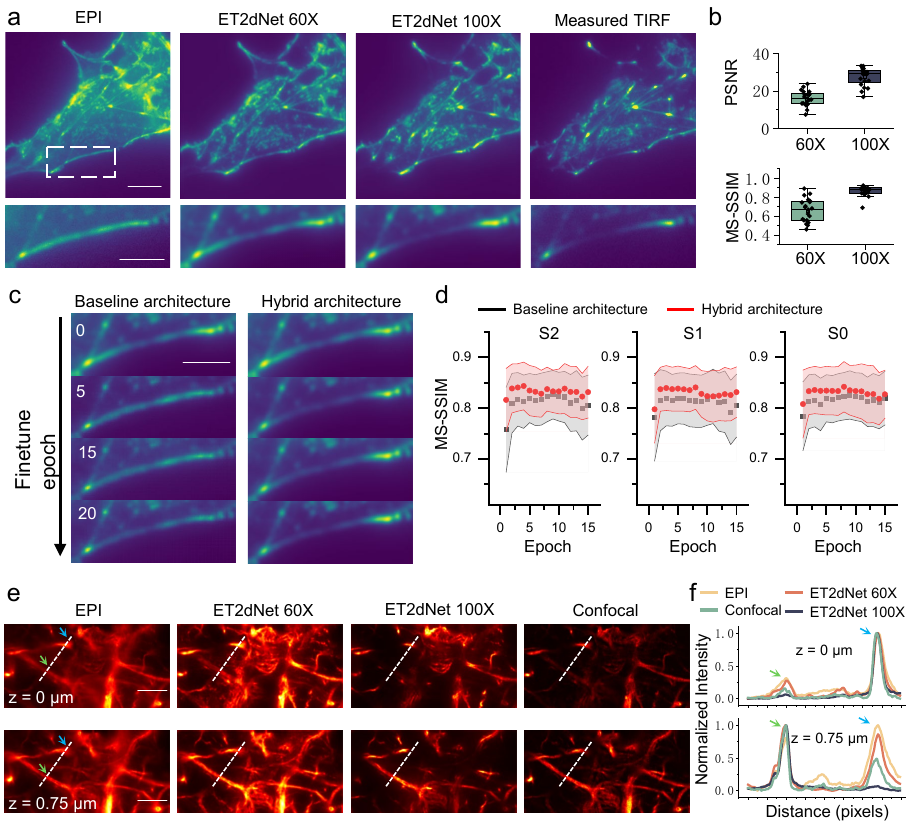}
\caption{\textbf{Few-shot adaptation of computational TIRF to new imaging systems} 
\textbf{a} Representative F-actin–labeled C6 cell images showing measured EPI (first column), cTIRF generated using a pretrained model trained on 60× data (ET2dNet 60X, second column), cTIRF after few-shot adaptation to a 100× imaging system (ET2dNet 100X, third column), and experimentally measured TIRF (fourth column). Magnified views of the regions indicated by white boxes are shown below.
\textbf{b} Quantitative comparison of cTIRF performance before and after few-shot adaptation, evaluated using MS-SSIM and PSNR across 20 paired EPI–TIRF image sets.
\textbf{c} Representative cTIRF outputs illustrating the adaptation dynamics of the baseline and hybrid network architectures during few-shot fine-tuning, shown at different adaptation stages for the region highlighted in (a).
\textbf{d} Statistical comparison of adaptation efficiency between the baseline (black) and hybrid (red) architectures, quantified by MS-SSIM during few-shot adaptation (mean $\pm$ s.d., smoothed over 3 epochs, 20 image pairs per condition).
\textbf{e} Representative images of a human brain tissue section stained for GFAP, showing measured EPI (first column), cTIRF generated using the pretrained model (second column), cTIRF after few-shot adaptation using the hybrid architecture (third column), and confocal microscopy (fourth column), at focal depths of 0 $\mu$m (top) and 0.75 $\mu$m (bottom). 
\textbf{f} Normalized intensity profiles along the line markers corresponding to (e).
Scale bar: 6 $\mu m$ in (e) and top of (a), 3 $\mu m$ in (c) and bottom of (a).
}\label{fig-finetunedET2d}
\end{figure}

Our deep learning method aimed to generalize across WF microscopes featuring diverse objectives (e.g., varying numerical aperture and magnification). We hypothesized a pretrained ET2dNet exhibited robust performance under similar PSFs (though non-identical), as validated in Fig.\ref{fig-ET2dNet}f. Here, input images acquired using a 60X/1.4NA objective were processed by a network trained on datasets acquired using a 60X/1.5NA objective. For systems exhibiting substantially different PSFs, fine-tuning was required. Implementation of the ET2dNet on WF images acquired with a 100X/1.5NA objective revealed effective background removal but negligible enhancement in axial-resolution, as shown in Fig.\ref{fig-finetunedET2d}a (denoted ET2dNet 60X). Specifically, at an arch-like structure (white dashed box), continuous microfilaments in the ET2dNet 60X output indicated a lack of axial-resolution improvement.

To improve the axial-resolution enhancement capability of source ET2dNet, we finetuned it on a small dataset of 20 paired EPI-TIRF images. The finetuning process achieved rapid convergence, yielding ET2dNet 100X, which significantly outperformed ET2dNet 60X at axial-resolution enhancement. For instance, at the arch-like structure indicated by the white box in Fig.\ref{fig-finetunedET2d} a, the ET2dNet 100X predicted discontinuous microfilaments, aligning with the measured TIRF. Quantitative validation across 20 test image pairs further confirmed improved PSNR and MS-SSIM after fine-tuning (Fig.\ref{fig-finetunedET2d}b). 

Furthermore, we demonstrated that the physics-informed hybrid network architecture enhanced generalization and advanced rapid adaptation. An ablation study investigating the impact of the self-supervised learning branch was presented in Fig.\ref{fig-finetunedET2d} c and d. Both the baseline architecture model (without self-supervised learning branch) and the hybrid architecture model were trained on the 60X dataset and subsequently fine-tuned on the 100X dataset. Visually, the hybrid model regained its axial-resolution enhancement capability within 5 epochs during fine-tuning, whereas the baseline model required more than 20 epochs, as shown in Fig. \ref{fig-finetunedET2d}c (the full images are provided in Supplementary Information Fig. S2). Statistical comparison in Fig. \ref{fig-finetunedET2d}d revealed the hybrid model initially achieved higher average MS-SSIM (to experimental measured TIRF in the test dataset) and converged faster to a superior final value, reinforcing our conclusion. Ablation experiments conducted on networks with different model sizes (represented by S2, S1, S0 in Fig. \ref{fig-finetunedET2d}d) consistently corroborated the above findings, underscoring the hybrid model's strong transferability and superior generalization capacity.

Axial resolution enhancement was further validated on images of stained human brain tissue sections (Fig. \ref{fig-finetunedET2d}e). Comparative analysis of ET2dNet implementations (60X vs 100X) on z-stack (containing two 0.75 µm axial spacing slices) revealed critical performance differences. While the 60X model provided partial background suppression, the fine-tuned 100X model substantially eliminated out-of-focus signals. Notably, two microfilaments (indicated by green and blue arrows) located at the 0 µm and 0.75 µm planes exhibited cross-plane interference in EPI images. As evidenced by intensity profiles (Fig. \ref{fig-finetunedET2d}e), both confocal microscopy and ET2dNet 60X failed to resolve these structures, whereas ET2dNet 100X successfully isolated the overlapping signals. This also demonstrated cTIRF can provide stronger optical sectioning than conventional confocal in our tested conditions.

\subsection{Volumetric computational TIRF reconstruction}
\subsubsection{training a 3D reconstruction network with ET2dNet and EPI z-stack}

While single-frame computational TIRF provides effective optical sectioning for focal planes, its extension to volumetric imaging introduces additional challenges along the axial dimension. In practice, volumetric reconstruction is typically performed from a finite stack, which can contain strong out-of-focus background from emitters located outside the reconstructed volume. In this setting, independently reconstructing each axial plane fails to enforce axial consistency and cannot jointly account for out-of-volume background that couples multiple planes. As a result, volumetric reconstructions can degrade in the presence of strong out-of-focus background (see examples Supplementary Information Fig. S3).
To address these challenges, we extend cTIRF to volumetric reconstruction by leveraging axial context across the stack, and implement this extension using a dedicated three-dimensional network (ET3dNet) trained via knowledge distillation\cite{wang_knowledge_2022} from ET2dNet .

\begin{figure}[htbp]
\centering
\includegraphics[width=0.9\textwidth]{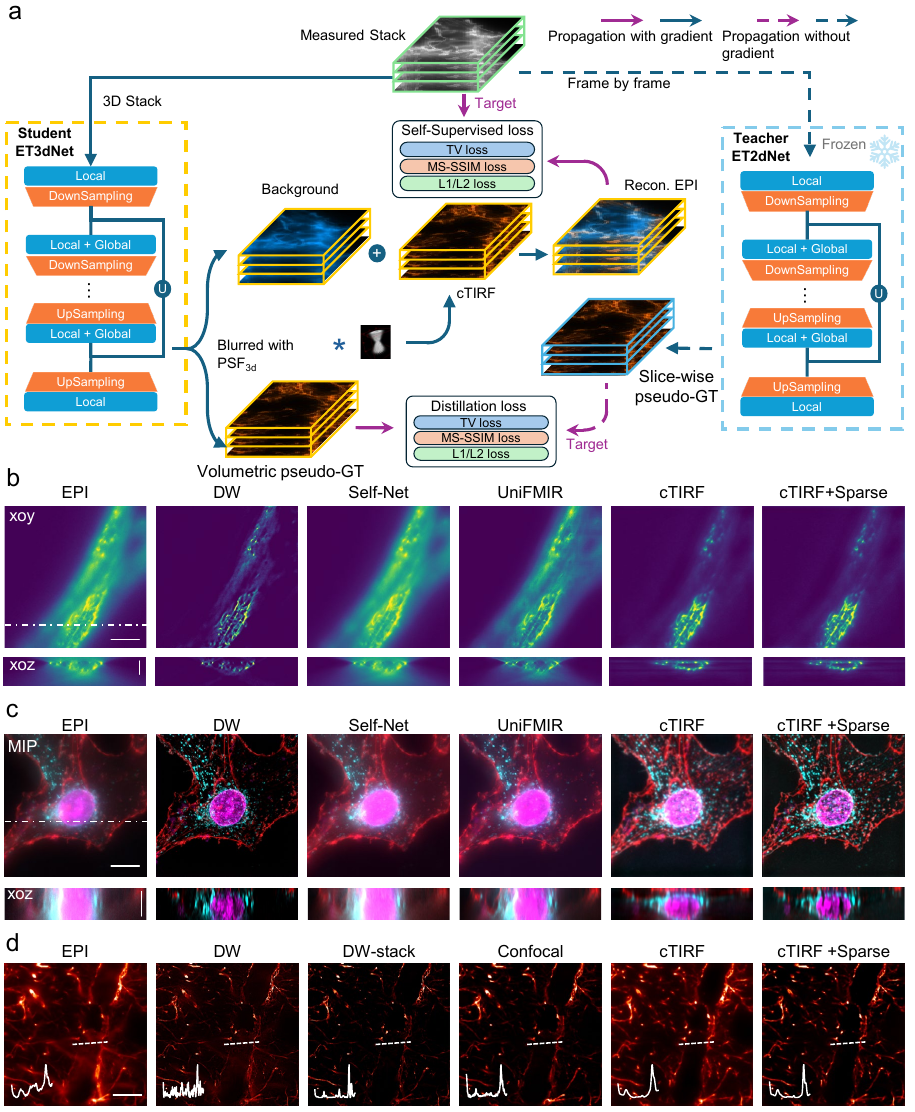}
\caption{\textbf{Characterizations and demonstrations of volumetric computational TIRF.} 
\textbf{a} The hybrid architecture of ET3dNet and the knowledge distillation process.
\textbf{b} Representative xy (top) and xz (bottom) slice of F-actin labeled C6 cell stack, acquired by EPI(first column), DW 3D reconstruction(second column), Self-Net (third column), UniFMIR(fourth column), volumetric cTIRF (ET3dNet outputs, fifth column), and Sparse Deconvolution reconstruction of volumetric cTIRF (sixth column).
\textbf{c} 
Three-color lateral maximum-intensity projections of a fixed C6 cell, maximum-intensity projections (top) and representative xz slice (bottom), acquired by EPI(first column), DW 3D reconstruction(second column), Self-Net (third column), UniFMIR(fourth column), volumetric cTIRF (fifth column), and Sparse Deconvolution reconstruction of volumetric cTIRF (sixth column). Red, F-actin labeled with Alexa Fluor™ 635 Phalloidin; blue, membrane labeled with Alexa Fluor™ 488 WGA; purple, nucleus stained by Hoechst 33258.
\textbf{d} Representative xy slice of human brain tissue stack containing 6-layers, acquired by EPI(first column), DW (second column), confocal(fourth column), volumetric cTIRF (fifth column), and Sparse Deconvolution reconstruction of volumetric cTIRF (sixth column). And DW of stacks containing 41-layers (third column).
Scale bar: 10 $\rm \mu m$.
}\label{fig-ET3dNet}
\end{figure}

ET3dNet was trained on 881 EPI image stacks (6 layers each) using the hybrid architecture depicted in Fig. \ref{fig-ET3dNet}a. Given the absence of 3D GT, we generated slice-wise pseudo-GT volumes using ET2dNet predictions from individual EPI layers. ET3dNet simultaneously outputs a volumetric pseudo-GT stack and a background stack. Supervised loss was computed between the volumetric pseudo-GT and slice-wise pseudo-GT. The volumetric pseudo-GT was convolved with the PSF (generates cTIRF stack) and combined with the background stack to generate reconstructed EPI stacks, enabling self-supervised loss calculation against experimental measured EPI stacks. The network backbone utilized EViT-Unet \cite{li_evit-unet_2025}. More details are shown in Methods \ref{sec-ETNet}.

\subsubsection{comparison of volumetric cTIRF and 3D reconstruction algorithm}

We assessed the performance of volumetric cTIRF, and made comparisons to SOTA 3D reconstruction algorithms (Fig.\ref{fig-ET3dNet} b, c, d), including self-supervised learning method Self-Net \cite{ning_deep_2023}, supervised learning method UniFMIR \cite{ma_pretraining_2024} and volumetric deconvolution method DW\cite{wernersson_deconwolf_2024}. Implementation details of DW, Self-Net and UniFMIR are provided in Supplementary Information Note 3. Implementations on F-actin-labeled C6 cells ( Fig.\ref{fig-ET3dNet} b) revealed that neither UniFMIR nor Self-Net achieved isotropic resolution. We supposed that UniFMIR's performance suffered from unrealistic synthetic training data. Self-Net requires clear lateral images, while the strong out-of-focus in such densely labeled sample deteriorates its performance. Although DW deconvolution outperformed learning-based methods, residual out-of-focus signal persisted in xz-planes (Fig. \ref{fig-ET3dNet}b). In contrast, volumetric cTIRF produced images with substantially reduced background and out-of-focus signals across both xy- and xz-planes. Notably, the volumetric cTIRF reconstruction exhibited lower lateral sharpness than the DW outputs. However, this limitation could be mitigated by integrating volumetric cTIRF with Sparse Deconvolution. For implementation details, refer to Supplementary Information Note 2.

In the comparative analysis on triple-labeled C6 cells (F-actin, membrane, nucleus; Fig. \ref{fig-ET3dNet}c), the volumetric cTIRF reconstructions preserved finer membrane structural details, confirming its superiority over learning-based methods. 
In xz-planes, volumetric cTIRF revealed distinct arc-shaped membrane curvatures, indicating significantly improved defocus suppression. Furthermore, volumetric cTIRF demonstrated effective compatibility with Sparse Deconvolution. Collectively, volumetric cTIRF outperformed SOTA learning-based techniques and was comparable to SOTA deconvolution method for axial-resolution enhancement capability.

Deconvolution-based volumetric reconstruction assumes that the recorded signal in a finite axial stack arises from emitters within the reconstructed volume. This assumption necessitates sufficient axial sampling to achieve accurate reconstruction (see Supplementary Information Note 4). This requirement, however, limits imaging speed and increases photobleaching. Volumetric cTIRF alleviates this constraint by enabling robust reconstruction from a limited sampled axial stack containing only six planes (Fig.\ref{fig-ET3dNet}d). Under such limited-sampling conditions, DW deconvolution failed to remove out-of-focus background, as evidenced by the marked degradation observed when comparing reconstructions from 6-layer and 41-layer stacks(Fig.\ref{fig-ET3dNet}d second and third columns). By reducing the dependency on sufficient axial sampling while maintaining effective background suppression, volumetric cTIRF enables fast and reliable three-dimensional imaging, making it particularly well suited for live-cell applications.

\section{Discussion}\label{sec-discussion}

In this study, we establish computational TIRF as a computational imaging modality that transforms widefield epifluorescence microscopy into a TIRF-like representations without hardware modification. The single-frame cTIRF overcomes the intrinsic depth limitation of hardware-based TIRF while achieving effective background suppression and axial super-resolution from single widefield images. The practical deployability of cTIRF is enabled by a hybrid learning strategy that integrates supervision from experimentally measured EPI–TIRF pairs with physics-informed self-consistency based on the optical PSF. This combination improves generalization across imaging conditions and facilitates rapid adaptation to new imaging systems with minimal additional data. Beyond single-frame imaging, cTIRF is further extended to volumetric reconstruction through a three-dimensional network trained via knowledge distillation, enabling artifact-reduced 3d reconstructions from limited sampled axial stacks.

The axial resolution of our current method can still be further improved. Here, the capability is fundamentally constrained by the quality of the TIRF images in our training dataset. Advances in TIRF-based technology, such as multi-angle \cite{fu_axial_2016} and two-wavelength \cite{stabley_real-time_2015} TIRF, can achieve nearly 20nm axial resolution. These methods could be incorporated into future versions. Beyond improving the training dataset, employing some effective alternative network architectures, such as RCAN \cite{chen_three-dimensional_2021}, DFCAN \cite{qiao_evaluation_2021}, and GAN \cite{wang_deep_2019}, may further improve the performance of cTIRF.

As with many deep learning methods, when applying our approach on a new imaging system, we recommend that users fine-tune the network in order to achieve optimal performance. This involves acquiring paired EPI-TIRF images on the user’s system and replacing the corresponding PSF in the training (fine-tuning) pipeline. Due to our hybrid network architecture, this fine-tuning may require only a small training set (e.g., 20 image pairs) and relatively few epochs (e.g., $<$20), as illustrated in Fig. \ref{fig-finetunedET2d}c,d. Note that fine-tuning is not strictly necessary. If the user’s primary goal is background removal rather than maximizing axial resolution, our pre-trained networks (ET2dNet 60X and ET2dNet 100X) still perform well for many common use cases.

Our work also contributes new insights to algorithm‐based super‐resolution. It is well recognized that task‐specific deep learning models often suffer from limited generalizability when applied to different restoration tasks \cite{ma_pretraining_2024,moor_foundation_2023}. One promising strategy to address this is pretraining networks to learn priors across diverse tasks, which can then benefit new restoration goals \cite{ma_pretraining_2024,moor_foundation_2023,liu_deep_2020}. In the cases of axial resolution‐enhanced restoration and related tasks, the EPI-TIRF cross-modality task serves as an excellent source of such priors. We demonstrated that these learned priors could be successfully transferred to yield improved performance, as evidenced by the knowledge distillation from ET2dNet to ET3dNet. This advantage could be extended to related applications, such as time-lapse super-resolution \cite{qiao_neural_2025}, through a similar knowledge distillation process that considering temporal consistency. Collectively, we believe our work will facilitate biological discovery while promoting novel developments in computational super-resolution algorithms.

\section{Methods}\label{sec-method}
\subsection{Microscope system}\label{sec-setup}
The experimental dataset was acquired using a custom-engineered epifluorescence microscope with integrated TIRF functionality provided by Wenzhou Jingdao Technology Co., Ltd, as shown in Extended Data Fig. 3. The illumination subsystem combines a 638 nm single-mode fiber laser (Shanghai Xilong Optoelectronics Technology FC-638-100-SM) for EPI-TIRF illumination, and an LED source (Jingdao LED-DC, 405/488/638 nm) for EPI illumination. The translation of EPI-TIRF illumination was achieved by a motorized translation stage (GMT CXN60150-S). The laser beam was collimated by lens 1 ( f = 120 mm) and focused to the BFP of the objectives (Olympus, UPLAPO60XOHR 60X / UPLAPO100XOHR 100X, NA=1.50) by lens 2 (f2 = 150mm). Optical path control was implemented through Chroma ZET series spectral filters, including an excitation filter (ZET405/488/561/640xv2) and emission filter (ZET405/488/561/640m-TRFv2), with fluorescence separation managed by a dichroic mirror (ZT405/488/561/640rpcv2). Image formation utilized a scientific CMOS camera (Andor Zyla 4.3 PLUS) paired with a 200 mm focal length tube lens (Thorlabs TTL200-S8).

\subsection{Sample preparation}
The rat glioma cell line C6 was cultured in complete culture medium containing 89\% v/v high-sugar Dalberg's modified Eagle's medium (DMEM with 4.5 g L-1 D-glucose), 10\% v/v fetal bovine serum, and 1\% v/v penicillin-streptomycin mixture in complete culture medium. The cells were cultured in a humidified atmosphere that contained 5\% $\rm{CO_2}$ at 37°C. When not in use, the cells were cryopreservation at -80°C.

C6 cells (2.0 mL, 1 × $10^5$ cells mL-1) were inoculated into 6-well plates fitted with slides (2 cm × 2 cm) and incubated overnight. The cell medium was aspirated and the cells were rinsed three times with PBS. The cells were pre-fixed for 15 min in a 4\% paraformaldehyde solution, followed by 20 min in a fixation buffer. The fixed cells were then processed sequentially as follows: rinsed three times with PBS (each rinse lasting 5 min), permeabilized for 5 min in 1× permeabilization buffer, and finally rinsed three times with PBS. Then the microfilaments of the cells were labeled with Alexa Fluor™ 635 Phalloidin for 30 min, the cell membrane was labeled with Alexa Fluor™ 488 WGA for 30 min, and the nucleus was labeled with Hoechst 33258 for 10 min at room temperature. The labeled cells were rinsed three times in PBS. Finally, the slide containing the cells was placed onto a glass slide that had a depression filled with PBS. Subsequently, the slide was sealed with nail polish to enable observation.

\subsection{Image Pre-processing}
\subsubsection{Registration}\label{sec-registration}
For the ET2dNet training dataset, the registration of in-focus EPI–TIRF image pairs was performed, where a rigid translation was applied to the EPI image to maximize the mutual information between image pairs \cite{maes_multimodality_1997}. The calculated translation distance for either the x or y direction was always less than 2 pixels. For the ET3dNet training dataset, the same rigid translation was performed for each slice in the EPI z-stack to ensure consistency between the two datasets.

The registration was also performed on EPI-confocal image pairs (in Fig.\ref{fig-ET2dNet}f, \ref{fig-finetunedET2d}e, and \ref{fig-ET3dNet}d), since they were acquired using two distinct microscopy systems, details are provided in Supplementary Information Note 5.

\subsubsection{Training dataset enhancement and normalization}\label{sec-normalization}

The EPI-TIRF images are cropped to a size of $1024 \times 1024$ from the center of raw images with a size of $2048 \times 2048$. Then during the training process, a $512 \times 512$ area is randomly chosen, with a 80\% probability for random rotation. 

Due to the differences in illumination intensity and uniformity between the EPI and TIRF modes, the absolute intensity of EPI-TIRF images was not comparable. Hence, normalizations were performed on each EPI and TIRF image, where the intensity for each pixel in an image was divided by a maximum value to be in [0,1]. In practice, the percentile normalization with clipping was performed on the training dataset, which is formulated as:
\begin{equation}\label{eq:norm}
norm(\boldsymbol{I},p_{high})=\min(1,\frac{\boldsymbol{I}}{perc(\boldsymbol{I},p_{high})})
\end{equation}
where $perc(\boldsymbol{I}, p)$ denotes the p-th percentile of the input image $\boldsymbol{I}$. $p_{high}$ is typically set to  98 during training and testing. A variance threshold-based analysis was conducted to ensure that the EPI-TIRF image pairs contained sufficient information. Note that out-of-focus images, which may appear during inference of an image stack, were not included in the training dataset, and this may result in unexpected outputs. It is evident that the variance of an in-focus image is greater than that of an out-of-focus image. The problem of out-of-focus images was alleviated by multiplying each TIRF image by its own variance.

\subsection{ETNet}\label{sec-ETNet}

\newcommand{\imgEPI}{\boldsymbol{x}}
\newcommand{\imgTIRF}{\boldsymbol{y}}
\newcommand{\bg}{\hat{\boldsymbol{b}}}
\newcommand{\gt}{\boldsymbol{g}}
\newcommand{\predTIRF}{\hat{\boldsymbol{y}}}
\newcommand{\predEPI}{\hat{\boldsymbol{x}}}
\newcommand{\predGT}{\hat{\gt}}
\newcommand{\psf}{psf}

Given an EPI-TIRF image pair, denoted as $\imgEPI$ (EPI) and $\imgTIRF$ (TIRF), the network output layer was designed to have twice the number of channels as the input, enabling separate reconstruction of the ground truth (GT) and background components. Specifically, $(\predGT, \bg) = f_\theta(\imgEPI)$, where $f_\theta(\cdot)$ is the trainable model, $\predGT$ is the pseudo-GT, and $\bg$ is the predicted background. Here and throughout, the hat symbol $\hat{}$ denotes network outputs.

For a single-layer EPI image $\imgEPI$, the output is a two-channel map. The cTIRF image is given by $\predTIRF = \predGT * \psf$, and the reconstructed EPI image is obtained as $\predEPI = \predTIRF + \bg$.

The overall loss function is defined as:
\begin{equation}
    \mathcal{L}_{loss} = \mathcal{L}(\predTIRF, \imgTIRF) + \mathcal{L}(\predEPI, \imgEPI) + \mathcal{R}
\end{equation}
\begin{equation}
    \mathcal{R} = \beta_1 \mathcal{L}_{tv}(\predGT) + \beta_2 \mathcal{L}_{tv}(\bg) + \beta_3 \mathcal{L}_{l1}(\predGT, 0)
\end{equation}
where $\mathcal{L}(\cdot, \cdot)$ denotes the loss between two images, $\imgTIRF$ is the TIRF image, $\imgEPI$ is the input EPI image, $\predTIRF$ is the cTIRF image, and $\predEPI$ is the reconstructed EPI image. Here, $\mathcal{L}(\predTIRF, \imgTIRF)$ serves as the supervised loss, $\mathcal{L}(\predEPI, \imgEPI)$ acts as the self-supervised loss, and $\mathcal{R}$ is the regularization term. Each loss term is formulated as a weighted sum of several components, including L1 loss, L2 loss, total variation (TV) loss, and multi-scale structural similarity (MS-SSIM\cite{Wang_2003}) loss:

\begin{equation}
   \mathcal{L}(a, b) = \alpha_1 \mathcal{L}_{l1}(a, b) + \alpha_2 \mathcal{L}_{l2}(a, b) + \alpha_3 \mathcal{L}_{MS-SSIM}(a, b)
\end{equation}

\begin{equation}
    \mathcal{L}_{l1}(a,b) = \| a-b \|_1
\end{equation}

\begin{equation}
    \mathcal{L}_{l2}(a,b) = \| a-b \|_2^2
\end{equation}

\begin{equation}
    \mathcal{L}_{tv}(a) = \sum_{i,j} \left[ (a_{i,j-1} - a_{i,j})^2 + (a_{i+1,j} - a_{i,j})^2 \right]
\end{equation}

\begin{equation}
    \mathcal{L}_{MS-SSIM}(a,b) = 1 - \text{MS-SSIM}(a, b)
\end{equation}

For ET3dNet, let the input EPI stack be denoted as $\imgEPI_{3d}$ (where the subscript $3d$ indicates a 3D stack), the predicted volumetric pseudo-GT stack as $\predGT_{3d}$, and the predicted background stack as $\bg_{3d}$. A 3D point spread function $\psf_{3d}$ is used for convolution, resulting in the cTIRF stack: $\predTIRF_{3d} = \predGT_{3d} * \psf_{3d}$. The reconstructed EPI stack is then given by $\predEPI_{3d} = \predTIRF_{3d} + \bg_{3d}$.

Since true GT stacks are unavailable, we leveraged the pre-trained ET2dNet (with fixed parameters) to process each slice of the 3D EPI stack, generating a slice-wise pseudo-GT stack $\boldsymbol{g}_{3d}^\dagger$. This slice-wise pseudo-GT serves as the supervisory target for ET3dNet training. For details on training a teacher (ET2dNet) for ET3dNet, refer to Supplementary Information Note 6.

The overall loss function for ET3dNet is defined as:
\begin{equation}
    \mathcal{L}_{loss} = \mathcal{L}(\predGT_{3d},\gt_{3d}^\dagger) + \mathcal{L}(\predEPI_{3d}, \imgEPI_{3d}) + \mathcal{R} 
\end{equation}
where $\mathcal{L}(\predGT_{3d},\boldsymbol{g}_{3d}^\dagger)$ denotes the supervised loss computed over the pseudo GT stacks generated by the teacher and student networks, $\mathcal{L}(\predEPI_{3d}, \imgEPI_{3d})$ is the self-supervised loss between the reconstructed and input EPI stacks, and $\mathcal{R}$ is the regularization term:
\begin{equation}
    \mathcal{R} = \beta_1 \mathcal{L}_{tv}(\predGT_{3d}) + \beta_2 \mathcal{L}_{tv}(\bg_{3d}) + \beta_3 \mathcal{L}_{l1}(\predGT_{3d}, 0)
\end{equation}
All loss components ($\mathcal{L}_{l1}$, $\mathcal{L}_{l2}$, $\mathcal{L}_{tv}$, and $\mathcal{L}_{MS-SSIM}$) are extended to 3D accordingly.

\subsection{Implementation details}
ET2dNet and ET3dNet were implemented in PyTorch with the EViT-UNet backbone. All experiments were performed on a single NVIDIA RTX 4090D GPU (24GB VRAM). Training ET2dNet took approximately 4 hours, while ET3dNet required about 12 hours. The batch size was set to 4 for ET2dNet and 2 for ET3dNet. The AdamW optimizer was used with an initial learning rate of $5 \times 10^{-4}$, scheduled by a warm-up cosine annealing strategy (5 warmup epochs, total 200 epochs). The best model was selected based on the lowest validation loss. Fine-tuning of ET2dNet followed the same training strategy and achieved optimal performance after 100 epochs, requiring about 15 minutes on a single NVIDIA RTX 4080.

For details on environment setup, inference with pretrained models, training, or fine-tuning, please refer to the ETNet GitHub repository (\codeURL).

\subsection{PSF usage and generation}\label{sec-psf}
We used the simulated PSFs generated by the PSF Generator Fiji plugin with Born \& Wolf model, while the parameters were adjusted accordingly in each case.

\subsection{Axial resolution estimation using decorrelation analysis}\label{sec-decorrelation}

Axial resolution was quantified using decorrelation analysis \cite{descloux_parameter-free_2019}, a single-image resolution estimator that determines the highest spatial frequency $k_c$ containing reliable signal and relates it to the effective resolution. Compared with other quantitative resolution assessment methods, such as Fourier ring correlation (FRC) analysis\cite{koho_fourier_2019,zhao_quantitatively_2023}, decorrelation analysis circumvents the need for an empirical threshold and can therefore be applied universally across different imaging modalities. 

For the analysis, an EPI stack of 201 slices was acquired with an axial sampling of 100 nm. Each XY plane comprised 512×512 pixels with a lateral pixel size of 97.5 nm. ET2dNet inference was applied independently to each XY slice to generate a corresponding cTIRF stack. Prior to inference, pixel intensities for each XY slice were normalized by dividing by $perc(\boldsymbol{I},p_{high})$ (where $p_{high}=98$, see Eq.\ref{eq:norm}), and were rescaled by the same factor after inference. 

The raw EPI and cTIRF stacks were then re-sliced into 512 XZ and 512 YZ cross-sections (each 201×512 pixels). Sectorial resolution estimation was performed on each cross-section using eight angular sectors, and the cutoff frequency $k_c$ from the fifth sector, corresponding to the axial direction, was extracted. The mean $k_c$  values for the raw EPI stack were 0.19 (XZ) and 0.21 (YZ), whereas the corresponding values for the cTIRF stack were 0.60 (XZ) and 0.67 (YZ). These measurements correspond to an approximately threefold improvement in axial resolution. Further details are provided in Supplementary Information Note 7.

\backmatter

\bmhead{Data availability}
The training datasets (including EPI-TIRF image pairs and EPI stacks) are uploaded to Figshare at \dataURL.

\bmhead{Code availability}
A pytorch version of ETNet is available on GitHub at \codeURL.

\bmhead{Acknowledgements}
This work was supported by the financial support from the Research Fund of Wenzhou Institute, UCAS (Grant No. WIUCASQD2021009, WIUCASQD2021037). The authors thank  Prof. Xin Zhou, Jianwei Shuai, and Yaogen Shu for providing computational resources.

\bmhead{Author contribution}
L. Dai and Y. Tang supervised the research. 
Q. Li, Y. Tang and L. Dai conceived and initiated this project. 
Q. Li and C. Lou developed the code, processed 
relevant imaging data, designed the detailed implementations.
Q. Li, C. Lou and L. Dai wrote the manuscript, with inputs from all authors.
B. Gong, X. Chen, H. Chen, B. Li, Y. Wang  and S. Yang built the EPI-TIRF system.
Y. Cheng, J. Wang, Q. Li and B. Gong prepared and imaged samples. 
All authors discussed the results and commented on the manuscript.

\bmhead{Competing Interests}
Q. Li, C. Lou, S.Yang, B. Li, Y. Wang, Y. Tang and L. Dai have a pending patent application on the presented framework. The remaining authors declare no competing interests.

\bigskip


\bibliography{ETNet-bibliography}

\end{document}